\documentclass[
reprint,
showpacs,
nofootinbib,
 amsmath,amssymb,
 aps,
pra,
]{revtex4-1}

\usepackage{graphicx}% Include figure files
\usepackage{dcolumn}% Align table columns on decimal point
\usepackage{bm}% bold math
\usepackage[colorlinks=true, citecolor=blue, urlcolor=blue ]{hyperref}% add hypertext capabilities

\newcommand{\bq}{\begin{equation}}
\newcommand{\eq}{\end{equation}}
\newcommand{\ba}{\begin{array}}
\newcommand{\ea}{\end{array}}
\newcommand{\bqa}{\begin{eqnarray}}
\newcommand{\eqa}{\end{eqnarray}}

\newcommand{\pal}{\partial}

\newcommand{\ran}{\rangle}
\newcommand{\lan}{\langle}
\newcommand{\W}{\widetilde{W}}

\begin{document}

\title{Entanglement and Discord of superposition of Greenberger-Horne-Zeilinger states}
\author{Preeti Parashar}
\email{parashar@isical.ac.in}
\author{Swapan Rana}
\email{swapanqic@gmail.com}
\affiliation{Physics and Applied Mathematics Unit, Indian Statistical Institute, 203 BT Road, Kolkata, India}
\begin{abstract}
We calculate the analytic expression for geometric measure of
entanglement for arbitrary superposition of two $N$-qubit canonical orthonormal
Greenberger-Horne-Zeilinger ($GHZ$) states and the same for two $W$ states. In course of characterizing all kinds
of nonclassical correlations, explicit formula for quantum discord (via relative entropy) for the former class of states has been presented. Contrary to the $GHZ$ state, the closest separable state to the $W$ state is not classical. Therefore, in this case the discord is different from the relative entropy of entanglement. We conjecture that the discord for the $N$-qubit $W$ state is $\log_2N$.
\end{abstract}
\date{\today}
\maketitle

\section{\label{sec:level1}Introduction}
Quantum entanglement has emerged as a key resource for quantum
computing, quantum communication and information related
processes. Characterization and quantification of multipartite
entanglement is an interesting challenging problem in the field of
quantum information and computation. There are several approaches
to deal with the issue and various measures have been proposed
(see e.g. \cite{HHHH9}, \cite{GT9} for nice reviews). Although the bipartite
case is well understood, there is no \emph{uniform} measure in the multipartite
scenario. For multipartite pure states, a natural and most well studied measure is the
Geometric measure of entanglement (GME)
\cite{WG3pra,WEGM4qic,Wei09,tcw,sud,lcax10}.
GME is a \emph{distance measure} and for an arbitrary pure state
$|\psi\ran$, it is usually defined as
$E_G(|\psi\ran)=1-\Lambda_{\psi}^2$, where
$\Lambda_{\psi}=\max |\lan\phi|\psi\ran|$ and the maximum is taken
over the set of all product states. The notion of GME has
also been extended to mixed states through the convex-roof
construction. Another well-known measure, the Groverian measure of entanglement
\cite{BNO02, SSB04} is exactly the same as GME, up
to a square operation. The Groverian measure originated
 from the modified quantum search algorithm \cite{BNO02} and is
defined through the success probability in the search using the
given state as the initial state: \bq\label{intg2}
G(|\psi\ran)=\sqrt{1-P_{max}(|\psi\ran)}\eq where
\bq\label{intpmax2}P_{max}(|\psi\ran)=\max_{|\phi\ran~\in\mbox{ \{
product states \}}}|\lan\phi|\psi\ran|^2\eq Through out this article
we shall consider the definition (\ref{intg2}) for GME.

Apart from the context of quantum search algorithm, GME has several other
operational interpretations, for example, in additivity of channel capacity \cite{wh},
local state discrimination \cite{hmmprl}, construction of optimal entanglement witness
\cite{WG3pra} etc. It has recently been employed to determine the usefulness of pure states
as universal resources for one way measurement based computation \cite{appgme1}
and to show that most of the entangled states are too entangled to be useful \cite{appgme2}. On the
other hand, GME has been applied in condensed matter physics to detect phase transitions and
for characterization of ground state properties of many body systems \cite{appcondphy}. It is also closely
related to other known measures such as relative entropy of entanglement (REE) and robustness
of entanglement \cite{appgme3}.

It is known that the entanglement property may change drastically under superposition.
For example \cite{popescu}, superposition of two product states may yield a singlet
state and, on the contrary, superposition of two Bell states may lead to a product state.
This has drawn attention recently and bounds on several entanglement measures like von Neumann Entropy
\cite{lpsprl06}, Negativity \cite{ofpra07}, Concurrence
\cite{yyspra07} and GME (see e.g.,\cite{acin, slc}) have been obtained. However,
most of these works pertain to bipartite systems only. In multipartite scenario, the most widely
studied pure states are $GHZ$ and $W$ states. This motivates us to consider these states.
In this work, we will explicitly calculate the analytic expression of GME for arbitrary
superposition of two $N$-qubit \emph{canonical} orthonormal $GHZ$ states (and also for $W$ states)
and show that they do not saturate the bound obtained in \cite{slc}.
As is clear from the definition, analytic calculation of GME for an arbitrary state
is reasonably difficult since it involves nonlinear maximization
process. However, recently it has been shown \cite{Wei09} that if
a pure state is permutationally invariant, then the calculation of GME
becomes simplified. In this case, the optimal product
state can be taken as tensor product of the same single system
thereby reducing the number of variables in the
maximization process.

On the other hand, recent investigations reveal that there exist
quantum correlations other than entanglement. One such is the quantum discord \cite{discord1},
which basically quantifies the total non-classical correlations in a quantum state. So, even an unentangled
(separable) state can exhibit nonclassical correlations. We would like to characterize all kinds of
nonclassical correlations for our states mentioned above. Though discord is usually studied in bipartite setting,
here, our aim is to explore multipartite states and hence we shall follow the  approach of Modi et. al.
\cite{kavmodiprl}. In this unified view, the quantifications are done by the relative entropy, which makes this
approach more challenging. Indeed optimization of relative entropy is known to be a difficult problem, e.g., there
does not exist a formula for the REE even for the simplest case of arbitrary
2-qubit (mixed) states \cite{eisert03}. Very recently, the reverse problem of finding the set of entangled states for which the chosen separable state is the closest one, has been solved \cite{gg10}. The necessary and sufficient condition given therein would help us in our analysis.

The organization of this article is as follows: In Sec \ref{ii}, we calculate
the explicit expressions for GME of superposition of two orthonormal $N$-qubit $GHZ$ states and also
of two $W$ states, and compare the results.  In Sec \ref{iii}, quantum discord is derived analytically for the
former class of states. Though we hoped to pursue the superposed $W$ states as well, unfortunately discord is not known even for
the single (usual) $W$ state. Hence, as a first step, we conjecture the discord for $N$-qubit $W$ states in Sec \ref{iv}. Finally we discuss some possible extensions and related issues in Sec \ref{v}.

\section{GME of arbitrary superposition of two $GHZ$ and $W$ states}\label{ii}
In this section we consider the superposition of two
$N$-qubit $GHZ$ states and derive its GME. An analogous analysis is done for $W$ states and the results are then compared.

\subsection{GME of arbitrary superposition of two $N$-qubit
$GHZ$ states} The full set of canonical orthonormal $N$-qubit GHZ
states is given by (up to an irrelevant global phase) \bq
\label{GG1}
|G_i^{\pm}\ran=\frac{|B_N(i)\ran\pm|B_N(2^N-1-i)\ran}{\sqrt{2}}
\eq where $i=0,1,\ldots,2^{N-1}-1$ and $|B_N(i)\ran=|i_1i_2\ldots
i_N\ran$ is the `binary representation of the decimal number $i$
in an $N$-bit string'. In a more convenient notation, (\ref{GG1}) can be written
as \bq \label{GG2} |G_i^{\pm}\ran=\frac{|0i_2i_3\ldots
i_N\ran\pm|1\bar{i}_2\bar{i}_3\ldots \bar{i}_N\ran}{\sqrt{2}}\eq
where $i_k=0,1\quad\forall k=2,3,\ldots,N$ and a bar over a bit
value indicates its logical negation.

The superposition of two general
$GHZ$ states is hard to analyze analytically and one has to take recourse to
numerical techniques. Hence, for simplicity and for the sake of analytic results, we
shall confine ourselves to states from this orthonormal set. Let us now consider an
arbitrary superposition of two orthonormal GHZ states $|G_i^{\pm}\ran$
and $|G_j^{\pm}\ran$ ($i\ne j$) as follows \bq\label{GG3}
|G_{ij}\ran=\cos\alpha|G_i^{\pm}\ran+\sin\alpha
e^{i\gamma}|G_j^{\pm}\ran\eq

First of all we note that the state in (\ref{GG3}) is not
Schmidt decomposable (SD)\footnote{It is obvious that for $|\psi\ran=$ Schmidt $\{\lambda_i\}$, $P_{max}(|\psi\ran)
=\max\{\lambda_i\}$ and $E(|\psi\ran)=-\sum\lambda_i\log\lambda_i$.}. From the definition, if $|\psi\ran$ is SD,
then \bq\label{schm} |\psi\ran=\sum\sqrt{\lambda_i}|ii\ldots i\ran:=\mbox{ Schmidt }\{\lambda_i\}.\eq
So, tracing out any subsystem would give a separable state and it is a necessary condition for SD.
However, in this case, tracing out the third qubit from the 3-qubit state $|G_{01}\ran=\cos\alpha|G_0^{+}\ran+\sin\alpha
e^{i\gamma}|G_1^{+}\ran$ gives an entangled state, which can be easily checked by the PPT criterion.

We also note that (\ref{GG3}) is not permutationally invariant and hence the theorem in \cite{Wei09}
is not directly applicable to it. Though apparently it looks that the phase
$\gamma$ can not be driven out, we will show that the GME is
independent of $\gamma$.

To calculate the GME, let us  write the state (\ref{GG3})
in the following convenient form  \bq\label{GG4}
|G_{ij}\ran=c_1|\psi_m\ran|\phi_n\ran+c_2|\psi_m\ran|\bar{\phi}_n\ran
+c_3|\bar{\psi}_m\ran|\phi_n\ran+c_4|\bar{\psi}_m\ran|\bar{\phi}_n\ran\eq
where $|\psi_m\ran$ is the collection of (qu)bits when the two
strings $|B_N(i)\ran$ and $|B_N(j)\ran$ agree and $|\phi_n\ran$ is
the collection when they disagree, $m+n=N$;
$c_1=\frac{\cos\alpha}{\sqrt{2}}=\pm c_4$, $c_2=
 \frac{e^{i\gamma}\sin\alpha}{\sqrt{2}}=\pm c_3$.

 Noting that the right hand side of (\ref{GG4}) can be written as
 $|\psi_m\ran(c_1|\phi_n\ran+c_2|\bar{\phi}_n\ran)+|\bar{\psi}_m
 \ran(c_3|\phi_n\ran+c_4|\bar{\phi}_n\ran)$, it follows that\footnote{If $m=1$, then one possible optimal product state
is $|\Phi\ran=\sqrt{2}(c_1|\psi_m\ran+c_3|\bar{\psi}_m\ran)|\phi_n\ran$
and it produces $P_{max}=\frac{1}{2}$. The case $n=1$ is similar.
If $m,n\ge2$, the possible optimal product states are
$|\psi_m\ran|\phi_n\ran$, $|\psi_m\ran|\bar{\phi}_n\ran$ etc.
}
 \bq\label{GGpmax1}
P_{max}(|G_{ij}\ran)=\left\{
\begin{array}{l}
\max\{\frac{\cos^2\alpha}{2},\frac{\sin^2\alpha}{2}\} \mbox{ if }m,n\ge2 \\
\frac{1}{2}\mbox{ if } m=1 \mbox{ or }n=1
\end{array}
\right.\eq

The only known bound for GME of superposition (Eq. (3) in \cite{slc}) gives $P_{max}\le\frac{1}{2}+cs$, where $c=\cos\alpha,~s=\sin\alpha$. Since for $cs\ne0$, $\max\{\frac{c^2}{2},\frac{s^2}{2}\}<\frac{1}{2}$, $P_{max}$ can not saturate this bound.

\subsection{Proof of the result (\ref{GGpmax1})}
By suitable local unitary (LU) operations, the state (\ref{GG4}) can be transformed to
 the following form  \bq\label{nGG1}
|G_{ij}\ran=\frac{c}{\sqrt{2}}(|0\ran_m|0\ran_n\pm |1\ran_m|1\ran_n)+
\frac{s}{\sqrt{2}}e^{i\gamma}(|0\ran_m|1\ran_n+|1\ran_m|0\ran_n).\eq

Clearly (\ref{nGG1}) remains invariant under permutation among the first $m$ parties
(and similarly for the rest $n$ parties) and hence we can assume the optimal
product state to be of the form\footnote{This can be easily seen by defining a new
state involving only the first $m$ parties via projecting out the rest. This
state is symmetric under permutation and hence the assertion follows from \cite{Wei09}.}
 \bq\label{nGG2}
|\Phi\ran=(\cos\theta_1|0\ran+e^{i\lambda_1}\sin\theta_1|1\ran)^{\otimes m}
\otimes (\cos\theta_2|0\ran+e^{i\lambda_2}\sin\theta_2|1\ran)^{\otimes n}\eq
with $0\le\theta_1,\theta_2\le\frac{\pi}{2}$ and $0\le \lambda_1,\lambda_2\le\pi$.

In order to get $P_{max}(|G_{ij}\ran)$, we have to maximize the quantity \bqa\label{nGG3}2Q=|\cos^m\theta_1
(c\cos^n\theta_2+se^{i(\gamma-n\lambda_2)}\sin^n\theta_2)+\nonumber\\e^{-im\lambda_1}\sin^m\theta_1
(\pm ce^{-in\lambda_2}\sin^n\theta_2+se^{i\gamma}\cos^n\theta_2)|^2\eqa

Now, applying $|z_1+z_2|\le|z_1|+|z_2|$ successively, we have \bqa\label{ngg1}&&|\cos^m\theta_1
(c\cos^n\theta_2+se^{i(\gamma-n\lambda_2)}\sin^n\theta_2)+\nonumber\\&&e^{-im\lambda_1}\sin^m\theta_1
(\pm ce^{-in\lambda_2}\sin^n\theta_2+se^{i\gamma}\cos^n\theta_2)|\nonumber\\
\le&& \cos^m\theta_1|(c\cos^n\theta_2+se^{i(\gamma-n\lambda_2)}\sin^n\theta_2)|\nonumber\\&+&\sin^m\theta_1
|(\pm ce^{-in\lambda_2}\sin^n\theta_2+se^{i\gamma}\cos^n\theta_2)|\nonumber\\
\le&& \cos^m\theta_1(c\cos^n\theta_2+s\sin^n\theta_2)\nonumber\\&+&\sin^m\theta_1
(c\sin^n\theta_2+s\cos^n\theta_2)\eqa

Note that \bqa\label{max1}\max_{0\le\theta_1,\theta_2\le\frac{\pi}{2}}&\left[\cos^2
\theta_1(c\cos\theta_2+s\sin\theta_2)\right.&\nonumber\\&\left.+\sin^2\theta_1
(c\sin\theta_2+s\cos\theta_2)\right]=1&\eqa
which occurs at $(\theta_1,\theta_2)=(0,\alpha),(\frac{\pi}{2},\frac{\pi}{2}-\alpha)$; additionally
at $(\mbox{ arbitrary }\theta_1,\frac{\pi}{4})$ if $\alpha=\frac{\pi}{4}$.

So taking $n=1$ (the case $m=1$ is similar by symmetry) in (\ref{ngg1}),
\bqa\label{max2} &&\cos^m\theta_1(c\cos\theta_2+s\sin\theta_2)+\sin^m\theta_1
(c\sin\theta_2+s\cos\theta_2)\nonumber\\&\le& \cos^2\theta_1(c\cos\theta_2+s\sin\theta_2)+\sin^2\theta_1
(c\sin\theta_2+s\cos\theta_2)\nonumber\\ &\le& 1\eqa
where, in the second line, we have used $\cos^m\theta\le\cos^2\theta,\quad
\sin^m\theta\le\sin^2\theta\quad \forall m\ge2$. Thus, if $n=1$ or $m=1$, we have from
(\ref{nGG3}), (\ref{max1}) and (\ref{max2}), $P_{max}(|G_{ij}\ran)=\frac{1}{2}$.

Similarly using \bqa\label{max3}\max_{0\le\theta_1,\theta_2\le\frac{\pi}{2}}&\left[\cos^2
\theta_1(c\cos^2\theta_2+s\sin^2\theta_2)+\sin^2\theta_1\right.&\nonumber\\&\left.
(c\sin^2\theta_2+s\cos^2\theta_2)\right]=\max\{c,s\},&\eqa it follows that
$P_{max}(|G_{ij}\ran)=\max\{\frac{c^2}{2},\frac{s^2}{2}\},$ if $m,n\ge2$.\hfill $\blacksquare$

\subsection{GME of arbitrary superposition of two $N$-qubit $W$
states} The GME of superposition of two 3-qubit $W$ states has
been presented in \cite{WG3pra}. Although the generalization to
$N$-qubit case is quite straightforward, still we wish to derive
it here for the sake of completeness and comparison with the case
of GHZ states. [Interestingly, it is shown that the 4-qubit case
is the easiest one, easier than even the 3-qubit case, and an
explicit formula of GME has been derived].

Let us consider an arbitrary superposition of the two $N$-qubit
$W$ states, namely,
$|W\ran=\frac{1}{\sqrt{N}}(|00\ldots1\ran+\ldots+|10\ldots0\ran)$
and
$|\W\ran=\frac{1}{\sqrt{N}}(|01\ldots1\ran+\ldots+|11\ldots0\ran)$
as follows \bq\label{ww1}
|W\W\ran=\cos\alpha|W\ran+e^{i\gamma}\sin\alpha|\W\ran\eq

We note that the LU transformation
$\{|0\ran,|1\ran\}\to\{|0\ran,e^{-\frac{i\gamma}{N-2}}|1\ran\}$
leads to an overall phase and hence the phase $\gamma$ becomes
redundant. So without loss of generality, we can assume $\gamma=0$
and therefore \bq\label{ww2}
|W\W\ran=\cos\alpha|W\ran+\sin\alpha|\W\ran,\quad
0\le\alpha\le\frac{\pi}{2}\eq

Since the state $|W\W\ran$ is permutationally invariant (with positive
coefficients), we can assume the nearest product state as
\cite{Wei09}
\bq\label{wwphi1}|\phi\ran=(\cos\theta|0\ran+\sin\theta|1\ran)^{\otimes
N},\quad 0\le\theta\le\frac{\pi}{2}\eq

The overlap is given by \bq\label{wwip1}
|\lan\phi|W\W\ran|^2=Q=N[s C^{N-1}S+c
CS^{N-1}]^2\eq where $C=\cos\theta,\quad S=\sin\theta $, $\quad
c=\cos\alpha,\quad s=\sin\alpha$.

The condition for maximum of $Q$ ($\frac{\pal Q}{\pal \theta}=0$)
becomes \bq\label{wwmax1}s
[t^N-(N-1)t^{N-2}]+c[(N-1)t^2-1]=0\eq and hence we have
\bq\label{wwpmax1} P_{max}(|W\W\ran)=\frac{Nt^2}{(1+t^2)^N}
[c+ st^{N-2} ]^2\eq where $t=\tan\theta$ ($>0$)
will be determined from the polynomial equation (\ref{wwmax1}). We
note that Eq.(\ref{wwmax1}) has only one positive root for
$N=3,4$ and has at most three positive roots for $N\ge5$. Hence
the GME can be calculated using numerical techniques of root
finding.

Particularly, for $N=4$, we have \bq\label{WWg4q}
t^2=\frac{3(c-s)+
\sqrt{9-14cs}}{2c}\eq which readily gives the
expression for GME. The graph of GME vs. $s$ for 4-qubit case is
exactly analogous to the 3-qubit case ( Fig.-1. in \cite{WG3pra})
and hence we are not reproducing it here.

It is worth mentioning that if we consider the superposition of
two other  (even LU equivalent) $W$ states, the GME will not be the
same.

\subsection{Comparison between GME of superposition of $GHZ$ and $W$ states (with respect to the results presented here)}
1.\quad For $N=3$, either $m=1$, or $n=1$ and
    hence \bq
    P_{max}(|G_{ij}\ran)=\frac{1}{2}=P_{max}(|G_i^{\pm}\ran)\eq
    Thus for the three-qubit GHZ states, the GME of superposition
    is independent of the superposition parameter $\alpha$ and the
    phase
    $\gamma$, whereas for the $W$ states, it is
    dependent on $\alpha$. Of course, the GME of superposition
    of $|G_i^+\ran$ and $|G_i^-\ran$ (which is $\min\{|\frac{\cos\alpha+\sin\alpha
    e^{i\gamma}}{\sqrt{2}}|,|\frac{\cos\alpha-\sin\alpha
    e^{i\gamma}}{\sqrt{2}}|\}$) depends on both $\alpha$ and
    $\gamma$.

    For arbitrary $N$, if $m=1$ or $n=1$ (i.e., if the Hamming
    distance between $|B_N(i)\ran$ and $|B_N(j)\ran$ be 1 or
    $N-1$), then $G(|G_{ij}\ran)=\frac{1}{\sqrt{2}}$.

2.\quad For $N=3$, we note that by superposing  two orthonormal $W$ states,
    we can get the resultant entanglement equal to that of a GHZ
    state. For example, $G(\frac{|W_1\ran+|W_2\ran}{\sqrt{2}})=\frac{1}
    {\sqrt{2}}=G(|G_i^{\pm}\ran)$, where $|W_1\ran=\frac{1}{\sqrt{3}}(|001\ran-\omega|010\ran+
    \omega^2|100\ran),|W_2\ran=\frac{1}{\sqrt{3}}(-|001\ran+\omega^2|010\ran-\omega|100\ran)$;
    $\omega$ being a complex cubic root of unity. If we consider the superposition of $W$ and
    $\W$, it follows from Fig-1 in \cite{WG3pra}
    that we can choose a specific value of $s$ to get $G(|W\W\ran)=\frac{1}{\sqrt{2}}=G(|GHZ\ran)$.
    On the other hand, we can not get the entanglement of a W state by
    superposing any two orthonormal GHZ states from the canonical
    set (\ref{GG1}) since $G(|G_{ij}\ran)\le\frac{1}{\sqrt{2}}<G(|W\ran)$.\\

    However, for $N=4$ the situation is different. By superposing two W states, we can get
    the entanglement of a GHZ state and vice-versa.
    For example, $G(\frac{|W_1\ran+|W_2\ran}{\sqrt{2}})=\frac{1}
    {\sqrt{2}}=G(|G_i^{\pm}\ran)$, where $|W_1\ran=\frac{1}{2}(|0001\ran+|0010\ran+
    |0100\ran+|1000\ran)$ and $|W_2\ran=\frac{1}{2}(|0001\ran+|0010\ran-
    |0100\ran-|1000\ran)$. It can be checked that by superposing
    $|G_0\ran$ and $|G_3\ran$ we can get the entanglement of
    the W state: $G(|G_{03}\ran)=\sqrt{\frac{37}{64}}=G(|W\ran)$, where
    $|G_{03}\ran$ is given by (\ref{GG3}) with $\alpha$ given by $\sin\alpha$
    (or $\cos\alpha$) $=\sqrt{\frac{27}{32}}$.\\

    For large $N$, the situation is somehow opposite
    to the 3-qubit case. Here we can always get the
    entanglement of a $W$ state by superposing two $GHZ$ state,
    but we don't know if the converse is also true. This requires
    further investigation.

3.\quad For $N=3$, even if we consider equal superposition
    ($\alpha=\frac{\pi}{4},\gamma=0$), the state
    $|G_{ij}\rangle$ is not invariant under permutation. The corresponding product
    state would also not be permutationally invariant in general.
    However, for some $|G_{ij}\rangle$, the optimal product state may still be
    permutationally invariant. As an example, the state
    $|G_{01}^+\ran=\frac{1}{2}(|000\ran+|111\ran+|001\ran+|110\ran)$,
    is not permutationally invariant. But we can choose the nearest
    product state as $(\frac{|0\ran+|1\ran}{\sqrt{2}})^{\otimes3}$. [Of course, there exist
    other optimal product states which are not permutationally invariant e.g.
    $|q\rangle|q\ran|+\ran$ where $q = 0, 1, -$ and $|\pm\ran = \frac{(|0\ran
    \pm|1\ran)}{\sqrt{2}}$]. This may lead to the intuition that
    the optimal product state may be permutationally invariant even if
    the state itself is not. However, in \cite{Wei09}, the authors have proved the stronger result that
    this is not the case if the state is genuinely entangled (or we look into those parties for which this state
    is entangled and symmetric, e.g., $|G_{01}^{+}\ran=|\phi^+\ran|+\ran$, so we have to consider only the first two particles.
    Being symmetric and entangled, the combined state of these two particles have the necessarily symmetric closest product states $|q\ran|q\ran$).

\section{Quantum Discord for the class of states (\ref{GG3})}\label{iii}
As mentioned earlier, we will follow the approach of \cite{kavmodiprl} to characterize and quantify all
kinds of correlations in a quantum state. The definitions of relevant quantities are:
\bqa\label{qddf1}\mbox{ Entanglement }E&=&\min_{\sigma\in \mathcal{D}}S(\rho\|\sigma)\\
\label{qddf2}\mbox{ Discord }D&=&\min_{\chi\in \mathcal{C}}S(\rho\|\chi)\\
\label{qddf3}\mbox{ Dissonance }Q&=&\min_{\chi\in \mathcal{C}}S(\sigma\|\chi)\\
\label{qddf4}\mbox{ Classical correlations }C &=&\min_{\pi\in \mathcal{P}}S(\rho\|\pi)\eqa
where $\mathcal{P}$ is the set of all product states (i.e., states of the form $\pi=\pi_1\otimes\pi_2\otimes\ldots\otimes\pi_N$),
$\mathcal{C}$ is the set of all classical states (i.e., states of the form $\chi=\sum_{\overrightarrow{k}}p_{\overrightarrow{k}}
|\overrightarrow{k}\ran\lan \overrightarrow{k}|$, with the local states $|k_n\ran$ spanning an orthonormal basis), $\mathcal{D}$ is the set of all separable states (i.e., states of the form
$\sigma=\sum_k p_k \pi_1^k\otimes\pi_2^k\otimes\ldots\otimes\pi_N^k$) and $S(x\|y)=\mbox{Tr}[x\log x-x\log y]$
is the relative entropy of $x$ with respect to $y$. We shall first find out the closest separable state (CSS) to the class of states (\ref{GG3}). Fortunately it turns out that the CSS is also a classical state, thereby implying $D=E$, $Q=0$.

Before proceeding to calculations, we recall that finding out the CSS is a challenging problem
\cite{eisert03, hkpra10}. To obtain the CSS to a multipartite state,
two interesting tools are available in the literature. The first one is a lower bound through the generalization of Plenio-Vedral
formula \cite{plenved01}:
\bq\label{pvfre} S(\rho_N||\sigma_N)\ge S(\rho_{N-1}||\sigma_{N-1})+S(\rho_{N-1})-S(\rho_N),\eq
where $\rho_N$ is any $N$-partite state and $\sigma_N$ is an $N$-separable state. So, for any $N$-qubit pure state $\rho_N$ we have the lower bound \bq\label{qdf1}E(\rho_N)\ge\max\{E(\rho_{N-1})+S(\rho_{N-1})\}\eq
where the maximum is taken over all possible bipartition of $N-1$ versus single qubit.

The second tool is due to Wei et. al. \cite{WEGM4qic}. For any $N$-partite pure state $|\psi\ran$,
it gives a lower bound on $E$ through $P_{max}(|\psi\ran)$:
\bq\label{tcweire} E(|\psi\ran\lan\psi|)\ge-\log_2P_{max}(|\psi\ran).\eq Since $E$ is
defined through minimization, if we can find a separable state $\sigma$ which saturates either of the bounds in
(\ref{qdf1}) and (\ref{tcweire}), then $\sigma$ will be the required CSS. Infact the later bound has been extensively used
to derive REE of symmetric Dicke states \cite{WEGM4qic} and even mixtures of them \cite{tcweree08}. But unfortunately,
these bounds are not saturated for the states in (\ref{GG3}). Indeed, the bound (\ref{tcweire}) is not saturated even for the simplest 2-qubit non-maximally Bell states (e.g., for $|\phi\ran=a|00\ran+b|11\ran$ with $|a|^2+|b|^2=1$, one has $P_{max}=\max\{|a|^2,|b|^2\}$
whereas $E(|\phi\ran)=-|a|^2\log|a|^2-|b|^2\log|b|^2=H(|a|^2)$ \cite{vedral97}). It is thus quite challenging to derive the CSS. The reverse problem (i.e.,  starting from a $\sigma$ on the boundary of $\mathcal{D}$, determining all entangled state $\rho$ for which $\sigma$ is the CSS) is also interesting and has been solved for the 2-qubit case \cite{si03r}, very recently for multiparty states
\cite{gg10}. We shall apply this multi-party criteria to derive the CSS. The criteria reads:

\emph{\textbf{Necessary and sufficient criteria for CSS }}\cite{gg10}: $\sigma\in\mathcal{D}$ is a CSS for an entangled state $\rho$ if and only if \bq\label{gourcr}\max_{\sigma'\in D}\mbox{ Tr }\sigma'L_\sigma(\rho)=1,\eq
 where the linear operator $L_{\sigma}$ is defined in the following way. Let the eigendecomposition of hermitian positive operator $\alpha$ be $\alpha=\mbox{diag}(a_1,a_2,\ldots,a_n)$. Then for any $\beta=[b_{ij}]_{i,j=1}^n$, $L_{\alpha}(\beta)$ is defined by\bq\label{gourlab}[L_{\alpha}(\beta)]_{kl}=\left\{\begin{array}{ll}
                                                    b_{kl}\frac{\ln a_k-\ln a_l}{a_k-a_l}, & \mbox{if }a_k\ne a_l \\
                                                    b_{kl}\frac{1}{a}, & \mbox{if }a_k=a_l=a
                                                   \end{array}\right.\eq

We shall now derive the CSS of our states. Since REE is invariant under LU and (\ref{GG3}) can be transformed to (\ref{nGG1}) by LU, we can consider REE of this state, without loss of generality. The state (\ref{nGG1}) has GME similar to the non-maximal Bell state. So we assume that it will have a similar REE also. Hence we take the CSS as \bq\label{eq1}
\sigma=\frac{c^2}{2}(|00\ran\lan00|+|11\ran\lan11|)+
\frac{s^2}{2}(|01\ran\lan01|+|10\ran\lan10|),\eq
where (and hereafter) we have dropped the suffixes $m$ and $n$.

\textbf{Proof}:\bqa\sigma&=&\mbox{diag}(\frac{c^2}{2},\frac{s^2}{2},\frac{s^2}{2},\frac{c^2}{2})\nonumber\\
\mbox{and }\rho&=&\frac{1}{2}\left(
     \begin{array}{cccc}
       c^2 & cse^{-i\gamma}& cse^{-i\gamma} & \pm c^2 \\
      cse^{i\gamma}& s^2 & s^2 & \pm cse^{i\gamma}\\
      cse^{i\gamma}& s^2 & s^2 & \pm cse^{i\gamma}\\
      \pm c^2 & \pm cse^{-i\gamma}& \pm cse^{-i\gamma} & c^2 \\
     \end{array}
   \right).\nonumber\eqa
Hence from the definition of $L_\sigma(\rho)$,
\bq L_\sigma(\rho)=\left(
     \begin{array}{cccc}
       1 & qe^{-i\gamma}& qe^{-i\gamma} & \pm1 \\
      qe^{i\gamma}& 1 & 1 & \pm qe^{i\gamma}\\
     qe^{i\gamma}& 1 & 1 & \pm qe^{i\gamma}\\
      \pm1 & \pm qe^{-i\gamma}&\pm qe^{-i\gamma} & 1 \\
     \end{array}
   \right)\nonumber\eq where $q=\frac{cs\ln\frac{c^2}{s^2}}{c^2-s^2}$. Note that $|q|\le1$.

Now let $\sigma'=\sum p_k|\phi_k\ran\lan\phi_k|$. Then
\bqa &&\mbox{ Tr }\sigma'L_\sigma(\rho)=\sum p_k\lan\phi_k|L_\sigma(\rho)|\phi_k\ran\nonumber\\
&=&\sum p_k[|\lan\phi_k|00\ran|^2+|\lan\phi_k|01\ran|^2+|\lan\phi_k|10\ran|^2+|\lan\phi_k|11\ran|^2\nonumber\\
&+&2Real(qe^{-i\gamma}\lan\phi_k|00\ran(\lan01|\phi_k\ran+\lan10|\phi_k\ran))\nonumber\\
&\pm& 2Real\lan\phi_k|00\ran\lan11|\phi_k\ran+2Real\lan\phi_k|01\ran\lan10|\phi_k\ran\nonumber\\
&\pm& 2Real(qe^{-i\gamma}\lan\phi_k|11\ran(\lan01|\phi_k\ran+\lan10|\phi_k\ran))]\nonumber\\
&\le&\sum p_k[|\lan\phi_k|00\ran|^2+|\lan\phi_k|01\ran|^2+|\lan\phi_k|10\ran|^2+|\lan\phi_k|11\ran|^2\nonumber\\
&+&(|\lan\phi_k|00\ran||\lan\phi_k|01\ran|+\ldots+|\lan\phi_k|10\ran||\lan\phi_k|11\ran|)]\nonumber\\
&=&\sum p_k[|\lan\phi_k|00\ran|+|\lan\phi_k|01\ran|+|\lan\phi_k|10\ran|+|\lan\phi_k|11\ran|]^2\nonumber\eqa

Since each $|\phi_k\ran$ is a product state, we have
$|\phi_k\ran=|\varphi_k\ran|\psi_k\ran$. So the last expression above can be written as

$\sum p_k[(|\lan{\varphi}_k|0\ran|+|\lan{\varphi}_k|1\ran|)(|\lan{\psi}_k|0\ran|+|\lan{\psi}_k|1\ran|)]^2\le1,$ since for any normalized product state $|\phi\ran$ (of $\ge$ 2-qubits), $|\lan\phi|0\ran|+|\lan\phi|1\ran|\le1$ (which can be seen from (\ref{max3})).\hfill $\blacksquare$

\begin{figure}
  \includegraphics[width=7.5cm]{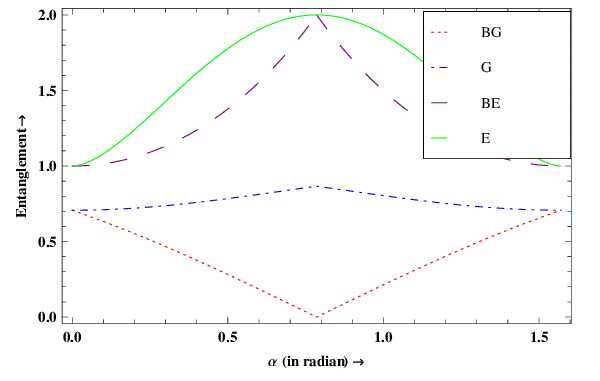}
  \caption{(color online only) Entanglement vs. $\alpha$ for the state (\ref{GG3}).  Exact results differ largely from the known bounds: BG, bound on GME obtained in \cite{slc}; G, exact GME; BE, bound on REE obtained through $-\log_2 P_{max}$ \cite{WEGM4qic}; E, exact REE.}\label{fig1}
\end{figure}

Thus $\sigma$ is indeed the CSS. Being a classical state as well, $\sigma$ is also the closest classical state (CCS), thereby yielding $D=E=-c^2\log\frac{c^2}{2}-(1-c^2)\log\frac{(1-c^2)}{2}
=1+H(c^2)$ and $Q=0$. We have depicted all the known bounds and our exact results for this state in Fig-\ref{fig1}.

\section{Conjecture for discord of $N$-qubit $W$ state}\label{iv}
From the discussion of the previous section, it is clear that determining CSS is a non trivial task. Determining the CCS is even more complicated because the set $\mathcal{C}$ is not a convex set and hence the standard tools of convex optimization theory is not directly applicable. However, to calculate the discord $D$ and dissonance $Q$, the authors of \cite{kavmodiprl} have simplified the task of minimizing over $\mathcal{C}$. They have shown that for any given $\rho$, the CCS $\chi_{\rho}$ is given by $\chi_{\rho}=\sum_{\overrightarrow{k}}|\overrightarrow{k}\ran\lan\overrightarrow{k}|\rho|\overrightarrow{k}\ran\lan\overrightarrow{k}|$, where $\{|\overrightarrow{k}\ran\}$ forms the eigenbasis of $\chi_\rho$. This simplifies expressions
for $D$ and $Q$ as the minimization of the relative entropy over $\mathcal{C}$ reduces to minimization of the
 von Neumann entropy $S(\chi_x)$ over the choice of local basis $\{|\overrightarrow{k}\ran\}$:
\bq\label{kavnedd} D=S(\chi_\rho)-S(\rho),\quad Q=S(\chi_\sigma)-S(\sigma),\eq
where $S(\chi_x)=\min_{|\overrightarrow{k}\ran}S(|\overrightarrow{k}\ran\lan\overrightarrow{k}|x|
\overrightarrow{k}\ran\lan\overrightarrow{k}|)$. Therefore, for numerical computation of $D$, one can choose arbitrary local bases and find the minimum of the corresponding entropies. An even finer approach is to generate a vector (with equal spacing) and using Gram-Schmidt method construct a complete orthonormal basis and obtain the minimum entropy. This technique is useful mostly in low dimensional cases \cite{nakahara}.

The CSS to the  $N$-qubit $W$-state is known to be \cite{WEGM4qic} \bq \sigma_W=\sum_{k=0}^N~^NC_k\left(\frac{k}{N}\right)^k\left(\frac{N-k}{N}\right)^{N-k}|S(N,k)\ran\lan S(N,k)|,\nonumber\eq
$|S(N,k)\ran$ being the $k$-th symmetric (Dicke) state. For $N\ge3$, the above separable state is not a classical state. Therefore $D\ne E$ and $Q\ne0$ for $W$ states (contrary to the $GHZ$ case, where the CSS was a classical state).

Since the $W$ state is symmetric, we assume that the CCS can be chosen to be symmetric\footnote{We assume it in the spirit of \cite{Wei09}. We note that in case of GME, the overlap function
(which has to be maximized) is a multilinear function of complex variables and so the optimization was easier because of availability of mathematical results. In case of REE, however, the entropy function is highly non linear and so far (to our knowledge) there is no such result for its optimization. If it can be proven true, the calculation of REE will be greatly simplified. However, even if it is not the case, still we hope our conjecture on $W$ state will hold, as it is supported by extensive numerical examples (this is why we make no comments on other permutationally invariant states). We note that the CSS is also symmetric.
}. So we choose each of the local orthonormal basis of the classical state $\chi_W$ as \bqa |0'\ran&=&\sqrt{p}|0\ran+\sqrt{1-p}|1\ran\nonumber\\
|1'\ran&=&\sqrt{1-p}|0\ran-\sqrt{p}|1\ran\nonumber\eqa so that $\lan x'|x\ran=(-1)^x\sqrt{p}$, $\lan
x'|y\ran=\sqrt{1-p}$, $x\ne y=0,1$. Therefore we have \bq\lan x'_1x'_2\ldots x'_N|y_1y_2\ldots y_N\ran=(-1)^{m_1}(\sqrt{p})^{m}(\sqrt{1-p})^{N-m},\nonumber\eq where $m$ is the number of positions where the two binary strings $x$ and $y$ agree and $m_1$ is the number of positions  where they both have 1. Since for the $W$ state each $y$ has exactly one 1, the inner product $\lan x' |W\ran$ will just depend on the number of 1's in $x$. So, if a basis $|x_k\ran$ has $k$ number of 1s, we have
\bqa\label{wdis1}\lan x'_k|W\ran=\frac{1}{\sqrt{N}}\left[-~^kC_1(\sqrt{p})^{N-k+1}(\sqrt{1-p})^{k-1}\right.\nonumber\\
+\left.~^{N-k}C_1(\sqrt{p})^{N-k-1}(\sqrt{1-p})^{k+1}\right]\nonumber\\
=\frac{1}{\sqrt{N}}(\sqrt{p})^{N-k-1}(\sqrt{1-p})^{k-1}[N(1-p)-k]\eqa

\begin{figure}
  \includegraphics[width=7.5cm]{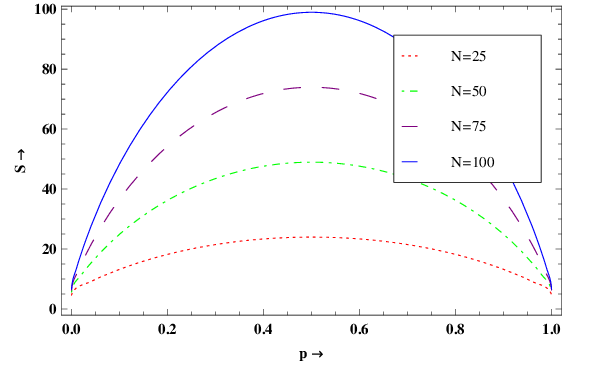}
  \caption{(color online only) $\mathbb{S}$ vs. $p$: minimum of $\mathbb{S}$ occurs at $p=0$ and $p=1$.}\label{fig2}
\end{figure}

Now from (\ref{kavnedd}), to determine $D$, we have to find the minimum of \bqa\label{wdis2}\mathbb{S}&=&S\left(\sum
\limits_{x'=x'_1x'_2\ldots x'_N;x_j=0,1}|x'\ran\lan x'|W\ran\lan W|x'\ran\lan x'|\right)\nonumber\\
&=&-\sum\limits_{x'=x'_1x'_2\ldots x'_N;x_j=0,1}|\lan x'|W\ran|^2\log_2|\lan x'|W\ran|^2\nonumber\\
&=&-\sum\limits_{k=0}^N~^NC_k\lambda_k\log_2\lambda_k,\eqa where $\lambda_k=|\lan x'_k|W\ran|^2$ with $x_k$ being any binary string of length $N$ having $k$ 1s, is given by (\ref{wdis1}). It can easily be checked that $\mathbb{S}$ has (global) minimum at $p=0,1$ (see Fig.-\ref{fig2}). Therefore the CCS to $W$ state is the dephased state in computational basis and consequently we have $D=\log_2N$.

Employing the method of \cite{nakahara}, we have also numerically verified (independent of  the assumption that the CCS is symmetric) that upto $N=5$, this indeed is the minimum. We thus \emph{\textbf{conjecture that discord of $N$-qubit W state is $\log_2N$}}.
\\

\section{Discussion }\label{v}
First  of all, we note that both the results (\ref{GGpmax1}) and (\ref{eq1}) can straightforwardly
be extended to the case of \textit{non-maximal} $GHZ$ states
(i.e., $a|i_1i_2\ldots i_N\ran+b|\bar{i}_1\bar{i}_2\ldots
\bar{i}_N\ran, |a|^2+|b|^2=1$). However, calculation of GME for
superposition of two \emph{arbitrary} GHZ states is more
involved. Fortunately in the present case we have been able to apply the result on permutationally invariant states \cite{Wei09}.
In fact, even for a single non-maximal (generalized) $W$
state, obtaining the GME is quite non-trivial. Recently, the three qubit
case has been studied in \cite{tcw} which has been further generalized
to $N$-qubits \cite{sud}. From a broader perspective, a generalization of GME in
which the maximum distance would be calculated from the set of all states which are equivalent
under stochastic local operations and classical communications (instead of just product states), has
been  introduced in \cite{ut}. It would be interesting to see how the GME of superposition behaves
in this context.

Another basic question related to the measure of correlations
 is the \emph{additivity} of the proposed measure. It is
known that GME is in general not additive \cite{wh}; precisely,
for $N\ge3$, GME is not additive for any two $N$-partite
antisymmetric states \cite{zch}. However, this is still not known for the case of total correlations $T_{\rho}$
( defined as $S(\rho\|\pi_{\rho})$) in a quantum state. It has been conjectured  \cite{kavmodiprl} that $T_{\rho}$ is
subadditive: $T_{\rho}>E+Q+C_{\sigma}$, where $C_{\sigma}$ is the \emph{classical correlation}
 $S({\chi}_{\sigma}\|{\pi}_{\sigma})$.

A further direction along our line of study would be to explore the correlations in $N$-qubit $GHZ$-diagonal states (an arbitrary mixture of the states $|G_{ij}^{\pm}\ran\lan G_{ij}^{\pm}|$). Because of the simple structure (both algebraic and geometric), the two qubit case allows easy computation of all the measures and  has been studied extensively. But, beyond this, even the criteria for entanglement is  unknown till date. We hope that the lower bound in (\ref{qdf1}) may provide some insight in determining the structure of the CSS which then can be verified using the necessary and sufficient condition given in \cite{gg10}.

To conclude, we have derived  analytically the GME and discord (via REE) for superposition of some orthonormal $GHZ$ states. We have also conjectured the discord for $W$ states. Perhaps a similar approach could be applied to other permutationally invariant states.

 \textbf{Acknowledgment}: SR would like to thank T.-C. Wei and Gilad Gour
for helpful discussions.

\end{document}